\documentclass[12pt]{iopart}

\usepackage{graphicx}
\begin{document}

\title[Q Wang et al]{A dedicated high resolution PET imager for plant sciences}

\author{Qiang Wang$^1$, Aswin J. Mathews$^2$, Ke Li$^2$, Jie Wen$^2$, Sergey Komarov$^1$, Joseph A. O'Sullivan$^2$ and Yuan-Chuan Tai$^1$}

\address{$^1$Department of Radiology, Washington University in St Louis, MO 63110, USA}
\address{$^2$Department of Electrical and Systems Engineering, Washington University in St
Louis, MO 63130, USA}
\ead{\mailto{wangqiang@mir.wustl.edu},\mailto{taiy@mir.wustl.edu}}

\begin{abstract}
PET provides in vivo molecular and functional imaging capability that is crucial to studying the interaction of plant with changing environments at the whole-plant level. We have developed a dedicated plant PET imager that features high spatial resolution, housed in a fully controlled environment provided by a plant growth chamber (PGC).
\\The system currently contains two types of detector modules: 84 microPET$^{\textregistered}$ R4 block detectors with 2.2 mm crystals to provide a large detecting area; and 32 Inveon$^{TM}$ block detectors with 1.5 mm crystals to provide higher spatial resolution. Outputs of the four microPET$^{\textregistered}$ block detectors in a modular housing are concatenated by a custom printed circuit board to match the output characteristics of an Inveon$^{TM}$ detector. All the detectors are read out by QuickSilver$^{TM}$ electronics. The detector modules are configured to full rings with a 15 cm diameter trans-axial field of view (FOV) for dynamic tomographic imaging of small plants. Potentially, the Inveon$^{TM}$ detectors can be reconfigured to quarter-rings to get a 25 cm FOV using step-and-shoot motion. The imager contains 2 linear stages to position detectors at different heights for multi-bed scanning, and 2 rotation stages to collect coincidence events from all angles. The detector modules and mechanical components of the imager are housed inside a PGC that regulates the environmental parameters.
\\The PET system has been built and integrated into the PGC. The system has a typical energy resolution of 15\% for Inveon$^{TM}$ blocks and 24\% for R4 blocks; timing resolution of 1.8 ns; and sensitivity of 1.3\%, 1.4\%, 3.0\% measured at the center of FOV, 5 cm off to R4 half-ring and 5 cm off to Inveon half-ring, respectively(with a 350-650 KeV energy window and 3.1 ns timing window). System spatial resolution is similar to that of commercial microPET$^{\textregistered}$ sytems, with 1.25 mm rod sources in the micro-Derenzo phantom resolved using ML-EM reconstruction algorithm. Preliminary imaging experiments using soybean and wild type and mutant maize labeled with $^{11}$C-CO$_{2}$ produced high-quality dynamic PET images that reveal the translocation and distribution patterns of photoassimilates.
\end{abstract}
\maketitle

\section{Introduction}
Rapid growing population brings in a fast increasing requirements on food, energy and other natural resources. Advances in molecular biology technique have made genetic modified corps that are more resistant to biotic(like insects, virus, microorganisms) and abiotic stress (such as drought, temperature extremes, nutrient limitation and so on) widely available to improve food and energy production. Practical corps yield and biomass growth is a complex phenotypic trait determined by the interactions of a genotype with the growth environment. Photonic-based techniques are widely applied in plant phenomics\cite{Yang2013}, X-ray computed tomography(CT) and MRI scanners are used to non-destructively image the inner structures of plant or its roots under soil\cite{Mooney2011}\cite{Borisjuk2012}. Besides the structural imaging tools, new physiological imaging methods is in great demands to rich the tool sets for future plant phenomics\cite{Fiorani2012}\cite{Dhondt2013}. Short-lived radioisotope technique can provide data that are crucial for developing more precise models that quantitatively link the underlying biochemical reactions to physiological response and better predictive models can be built to better understand the plant development and growth at the whole-body scope. This new method can also greatly save time and money for delivering new products to the farm compared with conventional field trails.
\\
PET is a noninvasive functional and molecular imaging technique that can provide quantitative information of dynamic radio-tracer distribution with spatial resolution in the order of millimeter. It is already commonly used to diagnosis human disease\cite{Phelps2000} and high spatial resolution scanners with small crystal size are also widely applied for small animal study\cite{Cherry2001}. 
Using PET detector modules to collect coincidence event counts of different part of plant is a conventional method at early time\cite{Minchin2003}. Then large planar detector modules with larger detection area are used to acquire projection images of interesting plant sections\cite{Kiser2008}. Commercial PET scanner has been used by some research groups for plant imaging\cite{Garbout2011}. Most of currently built PET scanners dedicated for plant imaging are based on these small crystal size PET detector modules which are usually adopted from small animal PET research projects. The Japanese group built a radio-tracer imaging system based on two large area planar detector head(120.8 x 186.8 mm$^{2}$) to acquire 2D dynamic projection PET images\cite{Uchida2004}. The detector head is composed of 4 x 6 detector modules each composed of 10 x 10 array of 2 x 2 x 20 mm$^{3}$ BGO scintillators. The German research group utilize 8 small animal PET detector modules developed by the ClearPET$^{TM}$ project\cite{Ziemons2005} to form two partial ring detector sets and achieve a 10.1 cm FOV. Completed data sets for 3D tomographic image reconstruction are acquired by rotational motion of the detector sets mounted on a rotation table\cite{Beer2010}. The plant PET scanner under development at Brookhaven National Lab is based on the RatCAP PET project\cite{Woody2004} with larger scanner bore size(100 mm in diameter and 18 mm in height) with more detector modules\cite{Budassi2012}. The proposed Jefferson National Lab's PhytoPET scanner features modular design concept to achieve re-configurable system geometry. The basic building module is a 48 x 48 array of 1.0 x 1.0 x 10 mm$^{3}$ LYSO crystals coupled to the 5 cm x 5 cm Hamamatsu H8500 position sensitive photomultiplier(PS-PMT)\cite{Weisenberger2013}.
\\
Unlike human or small animal imaging where the object size is somewhat fixed, the size of plants to be investigated may range from several millimeters to one meter, which mean the scanner should provide big FOV and high spacial resolution;
The temporal scale of the study may range from several minute using short half-life isotope($^{11}$C, $^{13}$N, $^{15}$O) to many days using long half-life isotopes($^{64}$Cu,$^{22}$Na), as a result, scanner should provide high sensitivity and low noise level; most positrons will escape from the thin leaf which needs special consideration\cite{Soret2007}\cite{Alexoff2011}\cite{Wu2011}; As plant are very sensitive to environment changes, controlled environment is crucial even at labeling and imaging time. And plants grow vertically, the plant PET bore axis must be vertical which is different from the human and small animal scanners.  To address issues specific to functional plant imaging, a dedicated plant PET imager has been built with the aim of providing a regional resource for plant sciences. 

\section{Material and methods}
\subsection{System design overview}
The proposed functional plant PET system is designed with two major features in mind: (1) high spacial resolution and sensitivity, potentially configurable system geometry to accommodate plants of different sizes and shapes and (2) the ability to control the environment in which the plants will be studied. To achieve these goals, we designed a PET system that is composed of high performance modular detectors for small animal PET scanners. Detector modules can be reconfigured to various geometry that are suitable for acquiring projection or tomographic images for different parts of a plant. These detectors are mounted to translation and rotation stages that are controlled by a computer remotely. The above components will be installed in a plant growth chamber with full environmental control. Radio-tracer delivery system has been developed to enable real-time radio-labeling and imaging capability for plant imaging studies.

\begin{figure}
\centering
\includegraphics[scale=.7]{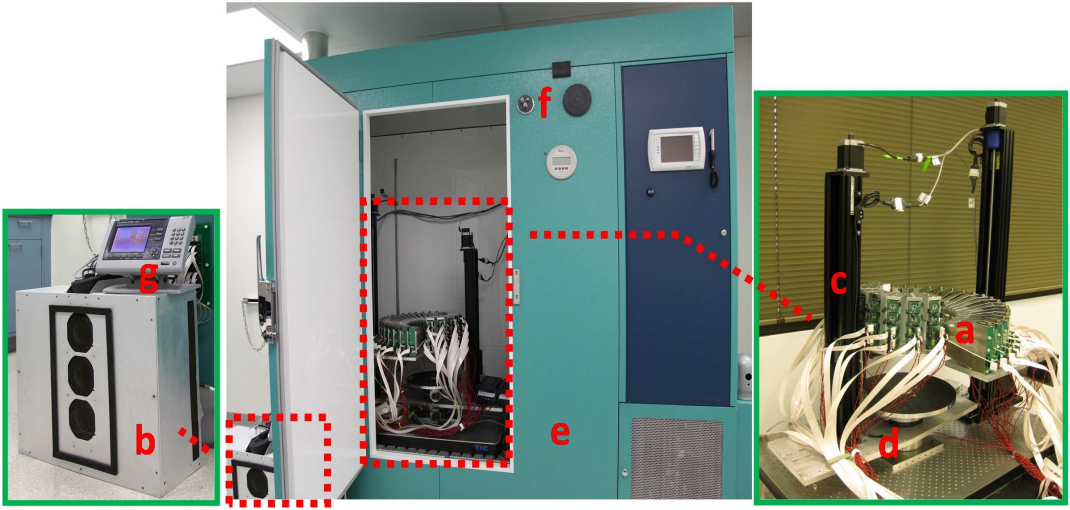}
\caption{Shows the plant PET system that is composed of (a) two sets of detector modules; (b) readout electronics system; (c) positioning stages; (d) mechanical support components; (e) plant growth chamber; (f) radio-tracer delivery port, and (g) motion monitoring and position indicators. Components a, c and d are inside the PGC that controls the imaging environment.}
\label{fig:SystemDiagram}
\end{figure}

\subsection{Detector modules and readout electronics}

\begin{figure}
\centering
\includegraphics[scale=.8]{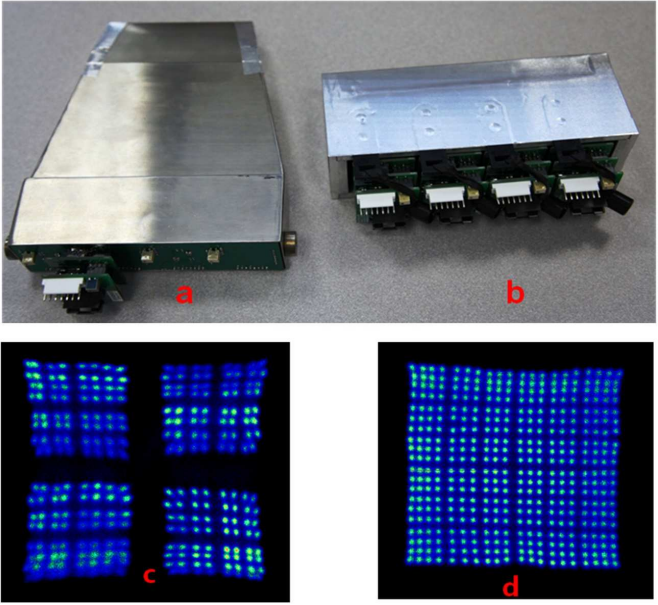}
\caption{(a) R4 detector module, (b) Inveon$^{TM}$ detector module, (c) The flood histogram of the whole R4 detector block. The outputs of four R4 block detectors are multiplexed by a custom PCB to re-map four 8x8 crystal arrays to form a 16 x 16 array to reduce the number of readout channels, (d) Inveon flood flood histogram.}
\label{fig:DetectorModule}
\end{figure}

We use two types of detector modules (shown in \fref{fig:DetectorModule}a,b) in the imager with different geometry to accommodate different imaging needs. The first group consist of 8 Siemens Inveon$^{TM}$ detector modules, each containing 4 PS-PMTs to readout 4 LSO crystal arrays. Each LSO array contains 20 x 20 crystals each measuring 1.51 x 1.51 x 10 mm$^3$ in 1.59 mm pitches. The second group consist of microPET$^{\textregistered}$ R4 detector modules, each containing 4 PS-PMTs to readout 4 LSO arrays. Each LSO array contains 8x8 crystals each measuring 2.2 x 2.2 x 10 mm$^3$ in 2.4 mm pitch. 21 R4 modules are used to build a detector panel with large solid angle coverage for imaging large plants. 
\begin{figure}[h]
\centering
\includegraphics[scale=.7]{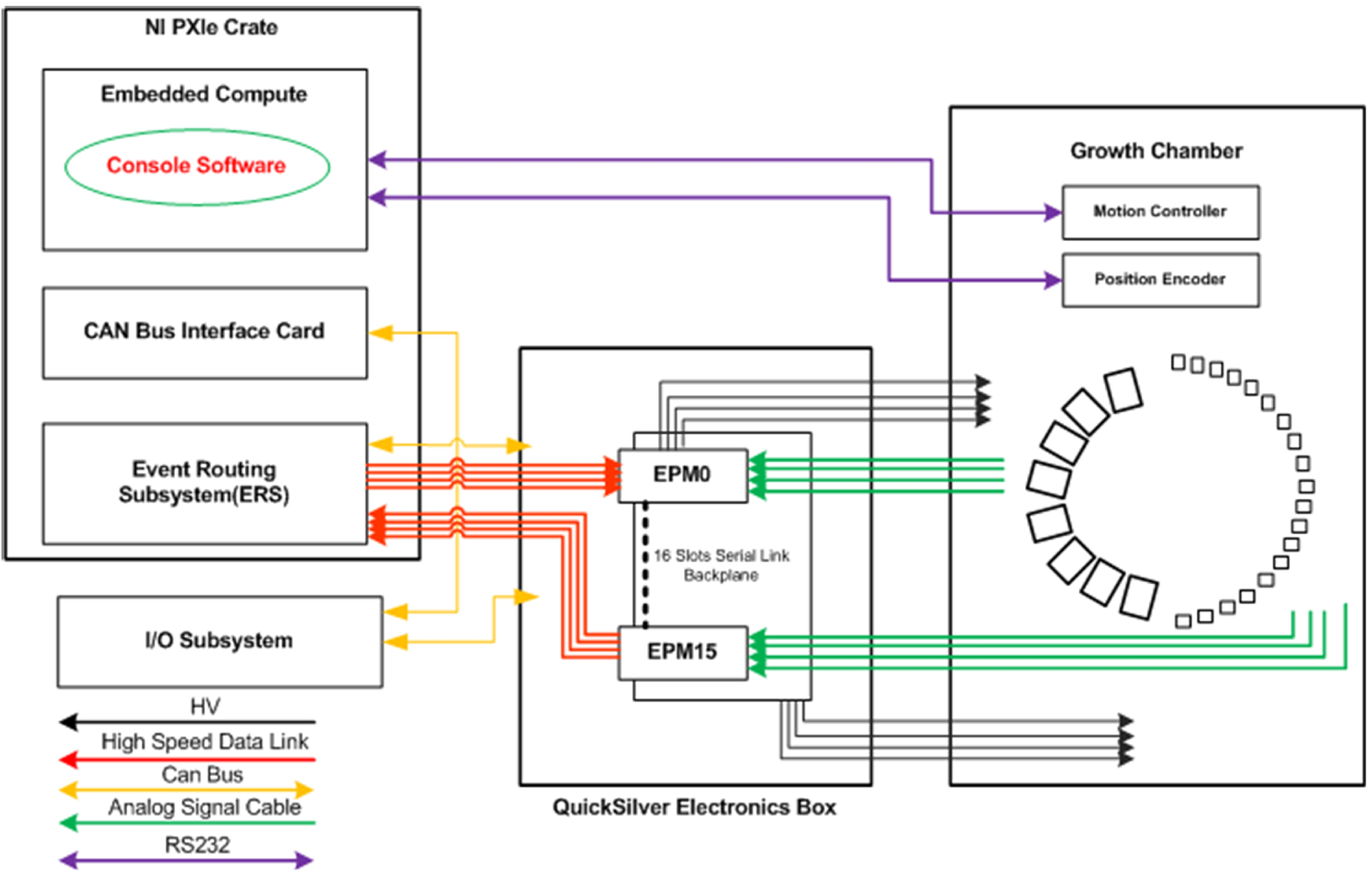}
\caption{Block diagrams of system readout electronics}
\label{fig:QuickSilverElectronics}
\end{figure}
Siemens QuickSilver$^{TM}$\cite{Newport2006} electronics is used for detector readout and coincidence processing. \fref{fig:QuickSilverElectronics} illustrates the data flow of the QuickSilver$^{TM}$ readout electronics system. The 16 output signals from 4 PS-PMT in each R4 module are multiplexed using a custom PCB to mimic a single PS-PMT with 4 position-encoded signals. This allow us to use the QucikSilver$^{TM}$ electronics to readout 32 Inveon$^{TM}$ detectors and 84 R4 detectors, and still have additional 11 readout channels for future silicon photomultiplier (SiPM) based sub-millimeter detectors that are under developed\cite{Song2010}. Two segments of flat flexible cables(total length of 3.5 m) are used to connect the detector output to the QuickSilver$^{TM}$ electronics through a custom designed junction board mounted on the PGC. \fref{fig:DetectorModule}c,d shows the flood histogram of a typical Inveon$^{TM}$ detector and a multiplexed R4 module read out by system electronics. No observable signal degradation is observed with these long cables.

\subsection{Reconfigurable geometry and positioning system}

\begin{figure}[h]
\centering
\includegraphics[scale=.8]{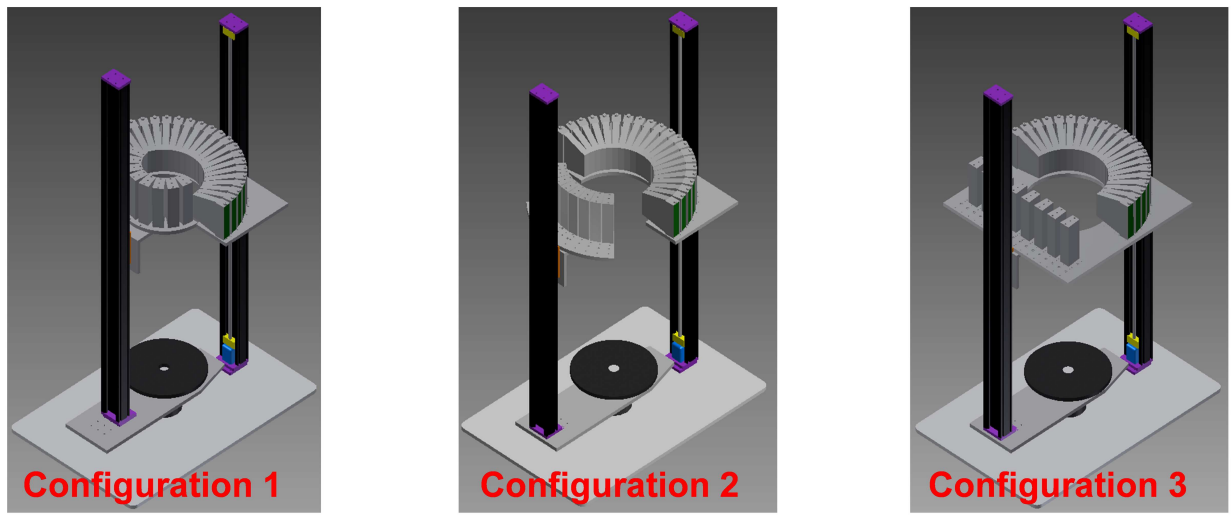}
\caption{Three different configurations of system geometry}
\label{fig:ReconfigurableGeometry}
\end{figure}

Eight Inveon$^{TM}$ detector modules (32 blocks) and  21 microPET$^{\textregistered}$ R4 detector modules (84 blocks) can be arranged to either form a half-ring (of different radii) to provide high-resolution dynamic imaging capability with an imaging FOV of 15 cm diameter by 10 cm tall(\fref{fig:ReconfigurableGeometry}.a). Alternatively, the Inveon$^{TM}$ modules can be arranged as a quarter-ring \fref{fig:ReconfigurableGeometry}.b. With a step-and-shot motion, this configuration provides tomographic images of objects up to 25 cm in diameter. With Configuration in \fref{fig:ReconfigurableGeometry}.c, the Inveon$^{TM}$ modules are arranged in a plane to get projection images of even larger objects.

\begin{table}[!ht]
\centering
\caption{\label{table:TableGeometry}Geometry parameters for different configurations}
\scalebox{0.8}{
\begin{tabular}{@{}c l l l l l}
\br
Config	&Actual FOV	&	&Imaging type	&Geometry/Radius (mm)&\\    
	uration& & & &
	\\
	&Axial	&Transverse	&	&Inveon	&microPET R4\\
	\mr
1	&10 cm *	&15 cm	&4D		&half-ring/86.1	&half-ring/140.7\\
2	&8 cm *	&25 cm	&3D or 4D ** &quarter-ring/166.6	&	\\
3	&12 cm *	&(up to) 40 cm	&2D projection	&planar/variable &	\\
\br
\end{tabular}
}
\\
\scriptsize
      Notes:\hfill\parbox[t]{14. cm}{* larger axial FOV (up to 60 cm) can be achieved with multi-bed scan\\
     *** depends on plant's traslocation speed} 
\end{table}

The motion of the detectors is controlled by the use of 2 linear stages and 2 rotation stages, mounted on a optical table (60 cm x 90 cm). As shown in \fref{fig:ReconfigurableGeometry}, the 2 vertical stages control the height of two sets of detector in order to track radiotracer throughout an entire plant. The radius of the R4 modules based half-ring is fixed while the distance between Inveon$^{TM}$ detector set to the center of the 2 concentric rotation stages can be adjusted by fixing the detector holding panel to different sets of holes drilled on the aluminum arm to accommodate plants of different size. A plant is typically centered at the top of ration stage. The quarter-ring detector group and the plant can be rotated independently to form lines of response(LOR) from all angles for tomographic imaging. Two rotation controllers are connected in a daisy-chain via a RS232 serial port to the host compute. The detailed geometry parameters of the three different configurations are shown in Table \ref{table:TableGeometry}. 

\subsection{Imaging console software}
\begin{figure}
\centering
\includegraphics[scale=3]{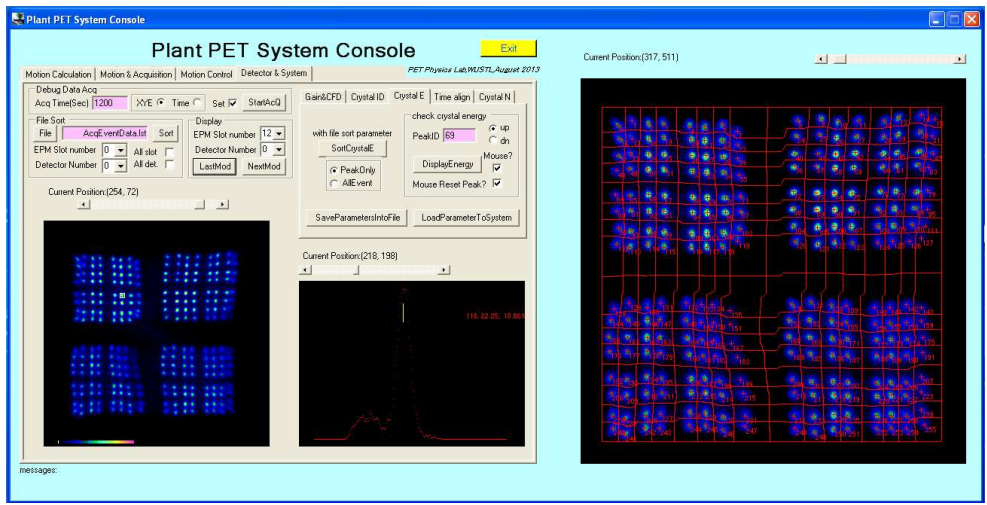}
\caption{Graphic User Interface (GUI) for detector and system setup. The screen snapshot shows an interactive crystal look-up table generating process}
\label{fig:ConsoleUI}
\end{figure}
The plant PET system have two different types of detector modules and it need some motion control function to image different size of plants with different detector geometry. A custom designed imaging console software (\fref{fig:ConsoleUI}) designed with MFC application framework provides the following main functions: (1) Detector modules and system setup; (2) Scanning angle and duration calculation based on system geometry and radio-nuclide half-life; and (3)Automated motion control an data acquisition.
\\Detector modules setup includes ASIC working parameters setup, crystal lookup table(CLU), energy lookup table(ELU), time alignment lookup table(TLU) generation. The supply voltages for Inveon and R4 modules are set to 700V and 800V individually. Tube gain difference of the four R4 detector blocks housed in the same module are mainly compensated by manually adjusting the high voltage divider resistors on the multiplexer boards. For Inveon$^{TM}$ modules, CLU, ELU and TLU are generated automatically. Setup of microPET$^{\textregistered}$ R4 modules requires minor manual effort to ensure correct identification of corner crystals. 
\\Automated motion control part calculates the needed motion steps and sends commands to the stepping motor controllers. The control codes also read back the position information from the decoders mounted on the stepping motors' axis to check if requested motions are completed.

\subsection{Image reconstruction}

The list-mode data are collected with QuickSilver$^{TM}$ electronics and sorted by custom sorting codes to sinogram data set. As multiplier boards are used for R4 detector modules, remapping codes are needed to convert the index ordered with flood histogram peaks to pre-defined reconstruction geometry coordinators. 
\\Reconstruction of PET images is based on the Maximum Likelihood estimation of activity concentration through the Expectation Maximization (ML-EM) algorithm. The system matrix is factored into a normalization component, attenuation component and geometric component. The geometric component of the matrix is computed by subdividing the detector crystals and forming sub-LORs joining the sub-crystals. Using Siddon's algorithm, the average intersection in each voxel and divided by the square of the length of the LOR to obtain the emission system matrix weights. Currently, as we do not have a method to estimate the attenuation of the subject under study, the attenuation component is ignored. For plants that have narrow stems and thin leaves, we suspect that the attenuation component is minor. The goal of normalization is to estimate the not modeled parameters in the system matrix. For this, we scan a Ge-68 phantom of known activity concentration for a period of 3 hours and estimate component through a maximum likelihood approach. The component efficiency that we estimate are the weights of R4-R4, R4-Inveon$^{TM}$,Inveon$^{TM}$-Inveon$^{TM}$ data and also, efficiency of individual crystals. Randoms and scatter estimates are added into the forward model. Random events rate is estimated through a delayed window approach. Scatter is estimated using a Single Scatter Simulation, whereby an image is reconstructed first, down-sampled, and scatter component estimated under a single scattering approximation. The tails of the scatter estimates are scaled and fitted to data.

\subsection{Growth chamber and radio-labeling system}

As plant is very sensitive to environment changes, plant growth environment needs to be controlled before and during imaging experiments. The plant PET imager is designed to be integrated in a plant growth chamber. The plant growth chamber (made by Conviron) has an exterior dimension of 79.5$^{\prime\prime}$ x 33.25$^{\prime\prime}$ x 79$^{\prime\prime}$ (WxDxH) and a interior 10 ft$^2$ growth area. The growth environment can be controlled with a temperature ranges from 4$^\circ$ Celsius to 45$^\circ$ Celsius, light intensity up to 1000 umoles/m$^2$/s, humidity level from 40\% to 90\%, and CO$_{2}$ level from the ambient level to above. 
The entire system is located in a plant-imaging lab (~24 m$^2$) above our cyclotron facility for easy access of a wide range of radionuclides and tracers. Gas isotopes are delivered via dedicated gas turnings directly from the cyclotron facility. The redundant radioactive gas or those flushing out from the labeling chamber can be recollected and delivered back to the cyclotron for further administration. Custom labeling chambers of different size or shape are made of conventional polyvinyl chloride(PCV) or transparent acrylic tubes. The radioactive gas are delivered either directly into the labeling chamber inside the PGC or to the radio-labeling system resided in a fume hood beside the PGC.

\section{Results}
\subsection{Basic performance measurements}

\begin{figure}
\centering
\includegraphics[scale=1.8]{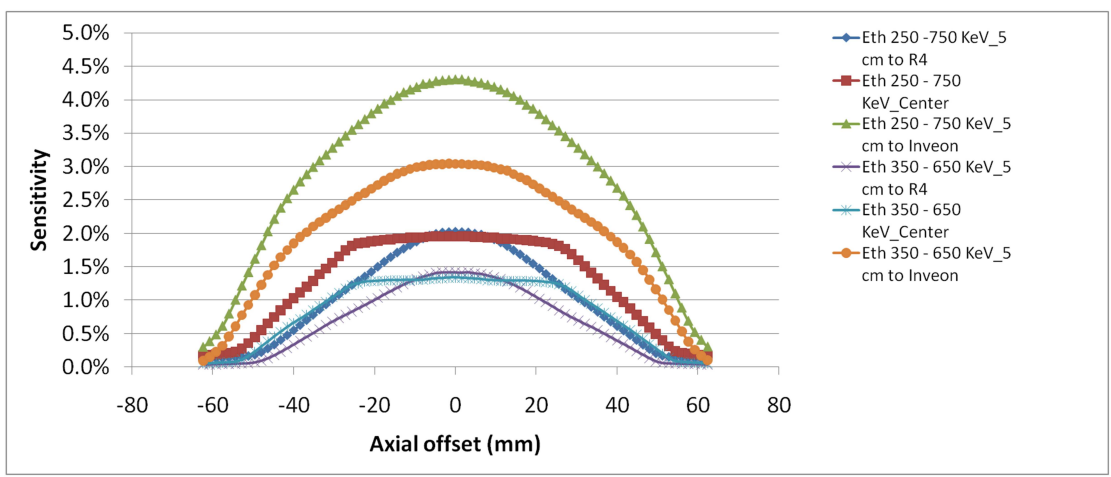}
\caption{The system sensitivity measure with different energy windows at 3 different positions crossing the axial FOV}
\label{fig:Sensitivity}
\end{figure}

The detector block energy resolution was measured using a Ge-68 line source, and was found to be 15.1+/-1.4\% and 23.8+/-6.2\% FWHM at 511 keV for Inveon$^{TM}$ and R4 detector, respectively. These results agree with thoese published in literatures\cite{Tai2001}\cite{Bao2009}. Coincidence timing resolution between an Inveon$^{TM}$ detector and a R4 detector was found to be 1.8 ns FWHM. The coincidence timing window of the system was set to 3.1 ns, accordingly.
For this particular detector geometry, system sensitivity was much different from conventional full ring systems. The sensitivity of the system shown in \fref{fig:Sensitivity} was roughly measured with a 70 uCi Ge-68 point source at 3 different trans-axial position (center, 5 cm off center to R4 half-ring, 5 cm off center to Inveon$^{TM}$ half-ring) crossing the whole axial FOV with a 1.6 mm step size. The system sensitivity at the axial center of those 3 selected positions is 1.3\%, 1.4\% and 3.0\% for energy window of 350-650 KeV, and 2.0\%, 2.0\%, 4.3\% for 250-750 KeV respectively with 3.1 ns time window. The sensitivity along the center axis within +/-25 cm offset maintains a peak value which may be useful for small plants study. The system sensitivity is limited by the solid angle coverage of the half-ring R4 modules. For the following imaging experiments, data was acquired using an energy window of 350 KeV to 650 KeV and a timing window of 3.1 ns.
\begin{figure}
\centering
\includegraphics[scale=0.8]{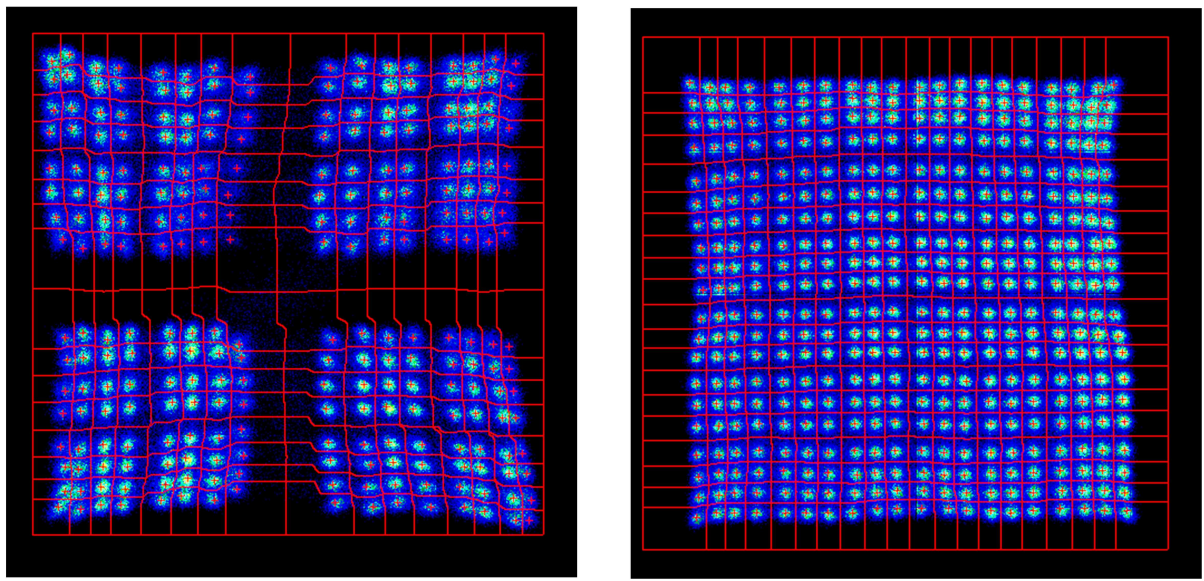}

\caption{The flood histograms of the two type of detector modules acquired at at 30$^\circ$ Celsius and the photon peaks and crystal lookup tables are generated based the data acquired at 20$^\circ$ Celsius (left:MicroPET R4 module, right:Inveon module). }
\label{fig:FloodTempEff}
\end{figure}
\\The plant PET scanner is located inside the PGC where the temperature may varies to mimic the change caused by the alternation of day and night. The LSO light output and the PMT's gain and quantum efficiency of the photo-cathode are affected by the environmental temperature\cite{Moszynski2006},\cite{Weber2003}. This variation may result in a shift of the photon peaks in the detector block's flood histogram or crystal's energy peaks. 25$^\circ$ Celsius and 30$^\circ$ Celsius are the common selected temperatures for day and night time respectively. A 70 uCi Ge-68 point source was used to acquire single events with 20$^\circ$ Celsius and 30$^\circ$ Celsius respectively. The detector modules was kept in the two temperature conditions for at lease 2 hours before taking data. The single events are sorted with our console program. \fref{fig:FloodTempEff} shows the crystals photon peaks found based on the data set acquired at 20$^\circ$ Celsius match very well with the same module's flood histogram acquired at 30$^\circ$ Celsius. The energy peaks of individual crystals from the two selected detector modules are compared and there a slightly decrease of energy peak value at 30$^\circ$ Celsius in the energy spectrum, but the energy resolution keeps the same. The evaluation shows that the detector module works stably with 10$^\circ$ Celsius of temperature variation which is meets the requirements from practical applications.

\subsection{Phantom study}
\subsubsection{Uniform phantom}
\begin{figure}
\centering
\includegraphics[scale=.8]{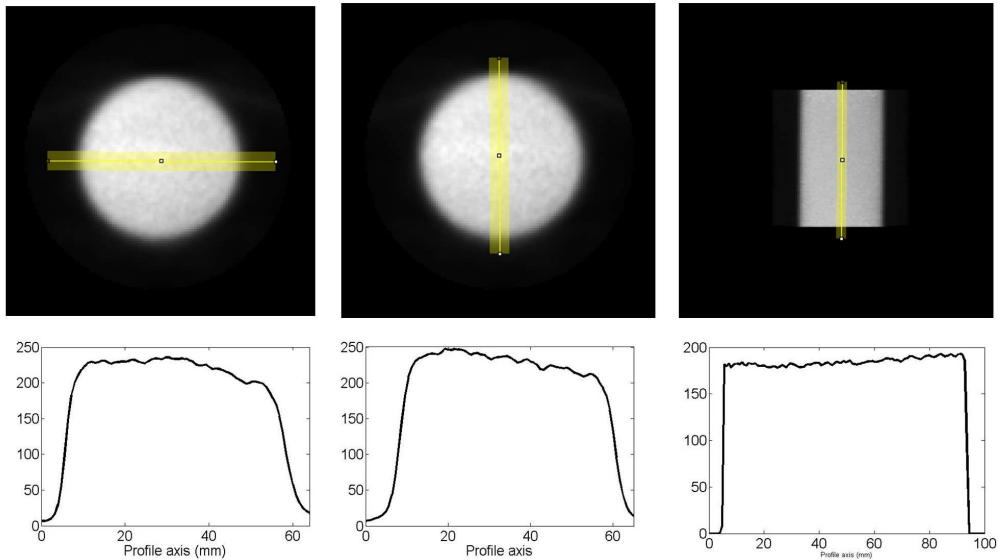}
\caption{Reconstructed uniform cylindrical phantom images and their profiles in tangential, radial and axial direction.}
\label{fig:UniformPhantom}
\end{figure}

A 6-cm diameter Ge-68 cylindrical phantom with a uniform activity concentration of 616 nCi/cc was used to normalize the system. A separate scan of the same phantom (with offset) was reconstructed with normalization and calculated attenuation correction. Images in \fref{fig:UniformPhantom} show good uniformity in the whole FOV.

\subsubsection{Derenzo-like pattern phantom}
\begin{figure}
\centering
\includegraphics[scale=1.0]{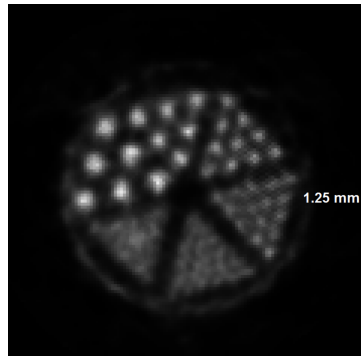}
\caption{Central slice of reconstructed micro-Derenzo phantom image with an inner core diameter of 3.2 cm and hot rod diameters of 0.80, 1.00, 1.25, 1.50, 2.00, 2.5 mm}
\label{fig:DerenzoPhantom}
\end{figure}

A home-made phantom with Derenzo-like hot rod pattern was scanned to evaluate the spatial resolution of the plant PET system. The inner core of the phantom has a diameter of 32 mm and contains fillable hot rods of different size (0.80, 1.00, 1.25, 1.50, 2.00 and 2.50 mm) arranged into 6 segments. The distance between adjacent rods in each segment is twice the rod diameter. The phantom was filled with 0.50 mCi(18.4MBq) of F-18 solution and scanned for 20 minutes. List-mode data was sorted into custom defined 3-dimensional sinograms and reconstructed with ML-EM algorithm. The 3-dimensional image size is 200 x 200 x 320 pixels with a 0.4 x 0.4 x 0.4 mm$^{3}$ voxel size. For this phantom study, a conventional energy window of 350 to 650 KeV and time window of 3.1 ns were applied. Phantom attenuation and scatter were not correction in reconstruction.

The reconstructed transverse slice of the phantom is show in \fref{fig:DerenzoPhantom}. The rods with 2.5, 2.0, 1.5 and 1.25 mm diameter are clearly separated and the 1.0 mm diameter group are moderately resolved. 
  
\subsection{Plant imaging experiments}
To evaluate the performance of this dedicated PET scanner for real plant imaging applications, three pilot experiments were conducted, which also provided the data of different plants' $^{11}$C-CO$_{2}$ absorption and translocation pattern. All of the following images are reconstructed with ML-EM algorithm and the image size is 400 x 400 x 160 pixels with a 0.8 x 0.8 x 0.8 mm$^{3}$ voxel size.

\subsubsection{First plant imaging experiment with cucumber plant}

\begin{figure}
\centering
\includegraphics[scale=.6]{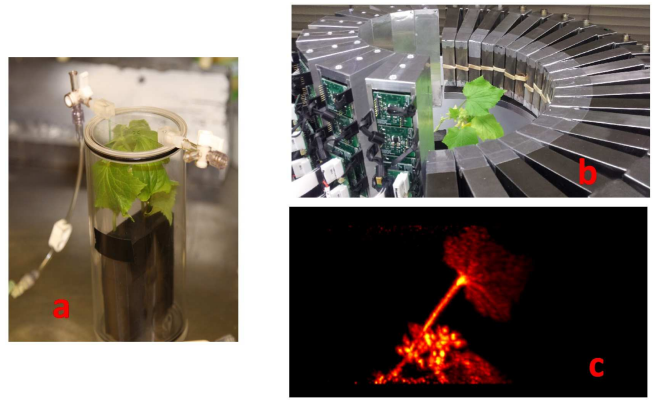}
\caption{The first PET imaging experiment with the system: (a) a cucumber plant being labeled with $^{11}$C-CO$_{2}$, (b) the plant is being imaged, (c) maximum intensity projection of the 3D PET images of the cucumber plant  clearly shows uptake in leaf, petiole, flowers and stem.}
\label{fig:CucumberImage}
\end{figure}

A young cucumber plant was labeled with 10 mCi $^{11}$C-CO$_{2}$ in a cylindrical chamber(shown in \fref{fig:CucumberImage}.a). The total uptake is ~0.3 mCi at the end of the 15-minute labeling. The plant was imaged for 10 minutes. Different parts of the cucumber plant are clearly delineated in the reconstructed image as shown in \fref{fig:CucumberImage}.c. The flowers appear to be the sinks of the photosynthates.



\subsubsection{Soybean imaging experiment}
\begin{figure}
\centering
\includegraphics[scale=1.0]{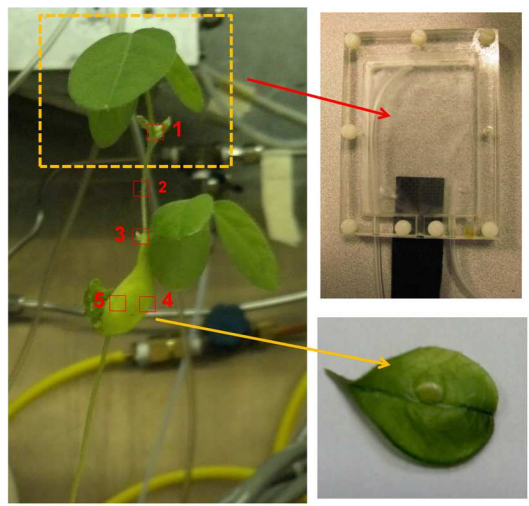}
\caption{spot labeling of top 3 leaves of a dwarf soybean plant}
\label{fig:SoybeanLabeling}
\end{figure}

The top 3 leaves of a dwarf soybean plant were labeled with 12 mCi $^{11}$C-CO$_{2}$ using a homemade rectangular labeling chamber(\fref{fig:SoybeanLabeling}) for 13 minute. The total uptake was estimated to be 6 mCi after decay correction back to the beginning of the labeling time. The plant was imaged 1 hour later (because it was too hot), beginning at time points 0, 60, 85 and 140 minutes, respectively.
\begin{figure}
\centering
\includegraphics[scale=.6]{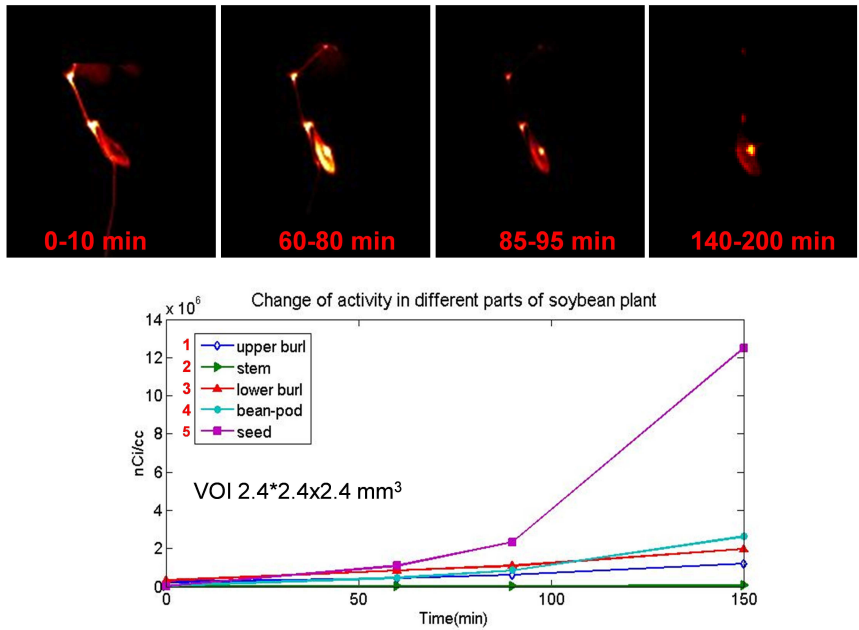}
\caption{reconstructed dynamic images at different time points (imaging started at 76 min post-injection), and time-activity-curve of selected VOIs}
\label{fig:DynamicSoybean}
\end{figure}

Five volumes of interest(VOIs) were selected at(1) the junction of leaves and stem, (2) stem, (3) junction of stem and a soybean pod, (4) edge of the pod, and (5) a bean inside the pod. Each VOI is 2.4 x 2.4 x 2.4 mm$^{3}$ (3 x 3 x 3 pixels in image). The mean value of the 27 voxels is plotted over time and shown as time-activity-curve of the VOI in \fref{fig:DynamicSoybean}. Most photosythates were translocated to the seed at the late frames.

\subsubsection{Maize root imaging experiment}
\begin{figure}
\centering
\includegraphics[scale=0.7]{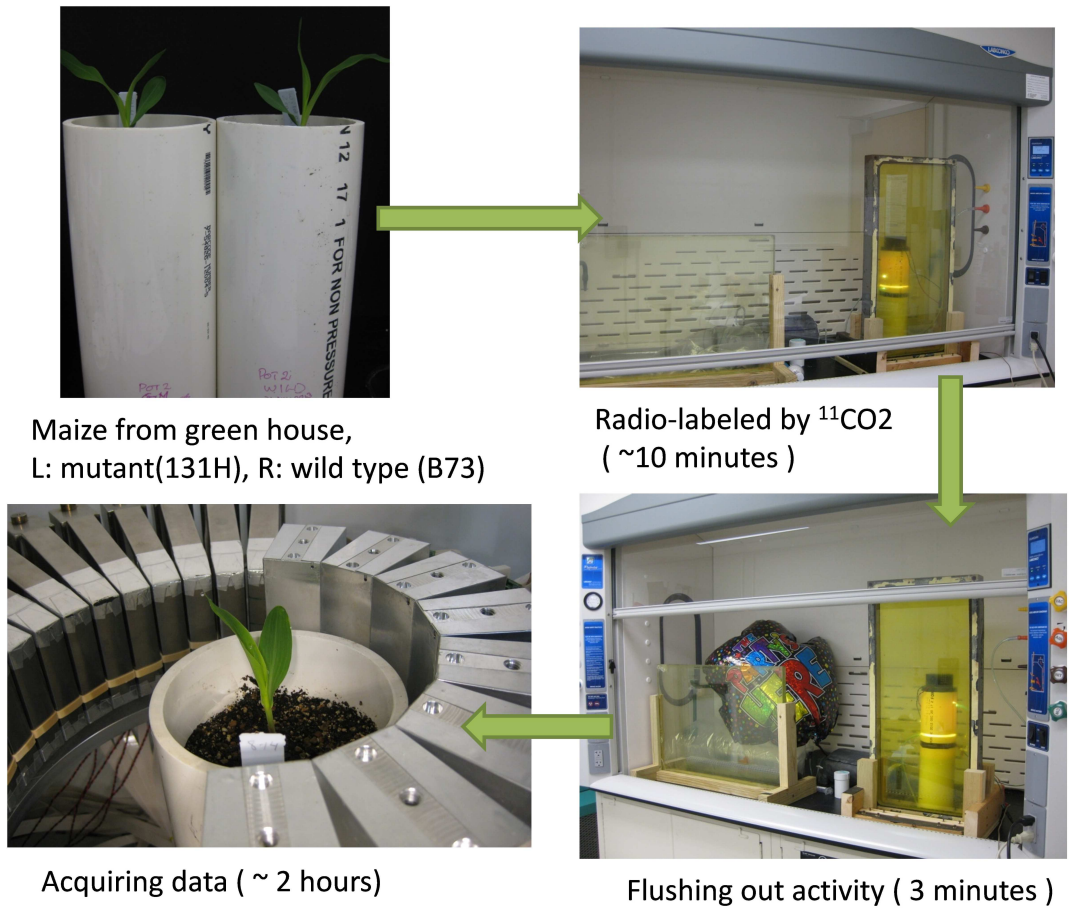}
\caption{shows the imaging protocol}
\label{fig:PlantLabeling}
\end{figure}

\begin{figure}
\centering
\includegraphics[scale=.8]{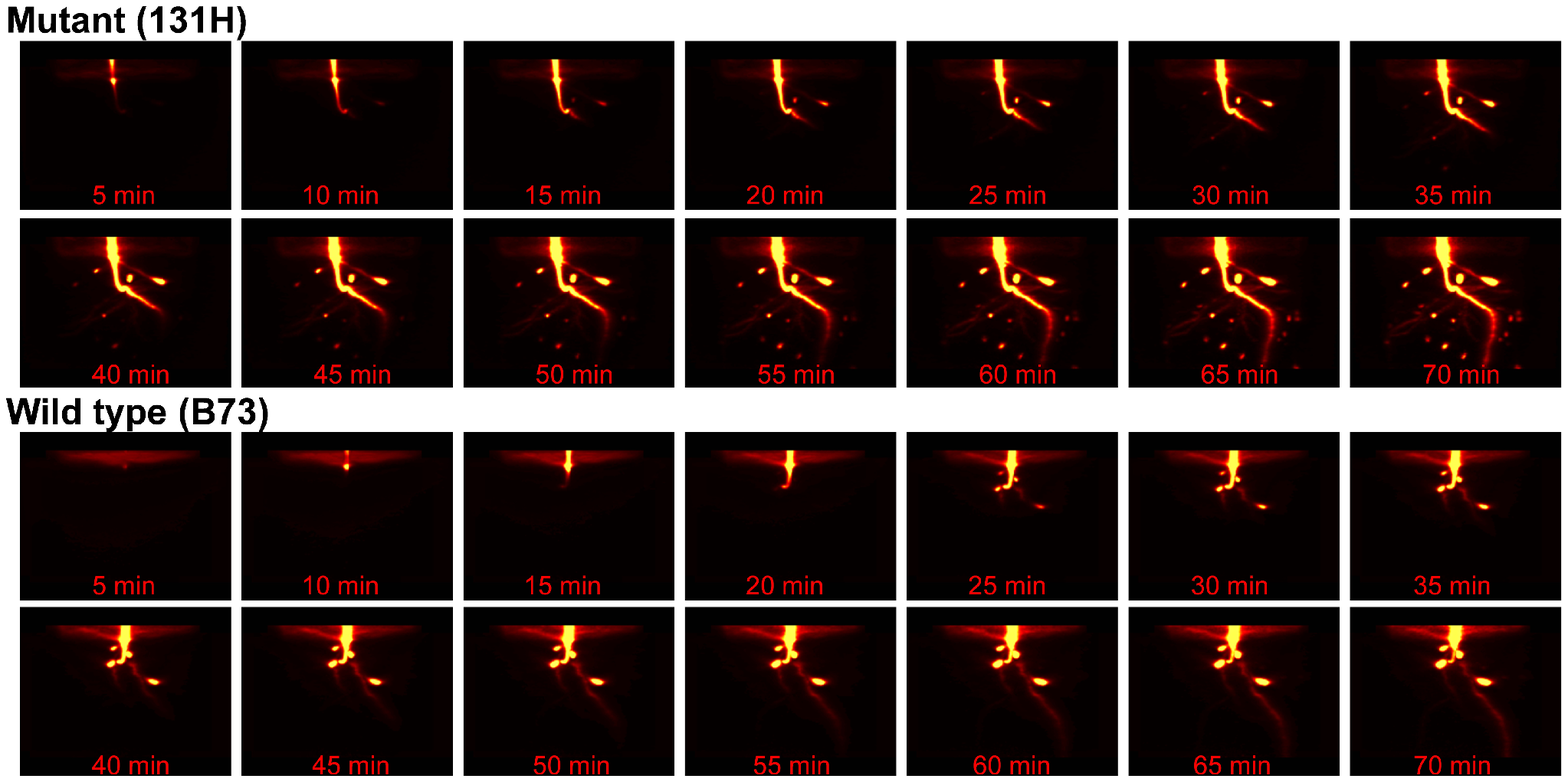}
\caption{Different translocation pattern shown with wild type and mutant maize dynamic PET image in 1 hour, upper: mutant, lower: wild type }
\label{fig:DynamicMutant}
\end{figure}

As shown in \fref{fig:PlantLabeling} and \fref{fig:DynamicMutant}, a young mutant and wild type maize (8 days after sowing) were labeled with about 10 mCi(370MBq) of $^{11}$C-CO$_{2}$ inside a custom made labeling chamber for about 10 minutes. The chamber was light up as soon as the $^{11}$C-CO$_{2}$ gas was injected by a high luminosity LED light source mounted on the chamber's top cap.  After the radio-labeling, the activity was flushed out and the plant was moved into the plant PET imager for scanning for about 2 hours. Raw data was binned into time frames with 5 minutes duration. No attenuation and scatter correction was applied in the following image reconstruction.
\\Dynamic images with 5-minute frames revel the different translocation and distribution patterns of photoassimilates in the 2 types of maize plants. The PET images at later time point (after 60 minutes) clearly show the root structures in regular soil. Small hot spots appeared at those small root ends, which possibly relates to a physiological truth that seeding root ends need more energy for new roots growth.    

\section{Discussion}
\subsection{Arbitrary geometry PET system}
This dedicated plant PET system features arbitrary geometry which practically reduce the total cost by reusing the detector modules from the old microPET$^{\textregistered}$ scanners. On the other hand, it provide more flexibility of detector modules can be used and detector geometry that would better fit our imaging objects, which is even more important for plant imaging. As mentioned above, the plant to be studied is of different size and shapes, a fixed detector geometry can not fulfill these requirements. Even the microPET$^{\textregistered}$ scanner that is dedicated for small animal study also has two different bore size to better target different imaging objects. Our study already shows that detector with different crystal size can work well in one readout system, and we have already test the signal compatibility of our high resolution MPPC based detector modules with the existed modules of PMT based. The reconstruction software also works with new add-in detector modules with different crystal size and geometry.    
    
Scanner's sensitivity is also a very important performance factor for plant imaging as the common used isotopes for plant labeling like $^{13}$N, $^{11}$C have very short half-life. The measured sensitivity at the center of FOV is almost the same when changing the setting from 4 Inveon$^{TM}$ detector half rings to 2 detector half rings. A third half-ring detector set can easily be added to the system to acquire coincidence events with the Inveon$^{TM}$ detector set and improve system sensitivity. The third detector set can be mounted apart from the existed R4 detectors, as a result,  we can simultaneously acquire coincidence events from two separated FOV with adjustable distance. This will be very useful for tall plant study, as the most important two parts of a plant are the top where flowers and young leaves are growing out and the bottom part with roots in soil. 
\subsection{Multi-bed scanning}
\begin{figure}
\centering
\includegraphics[scale=.7]{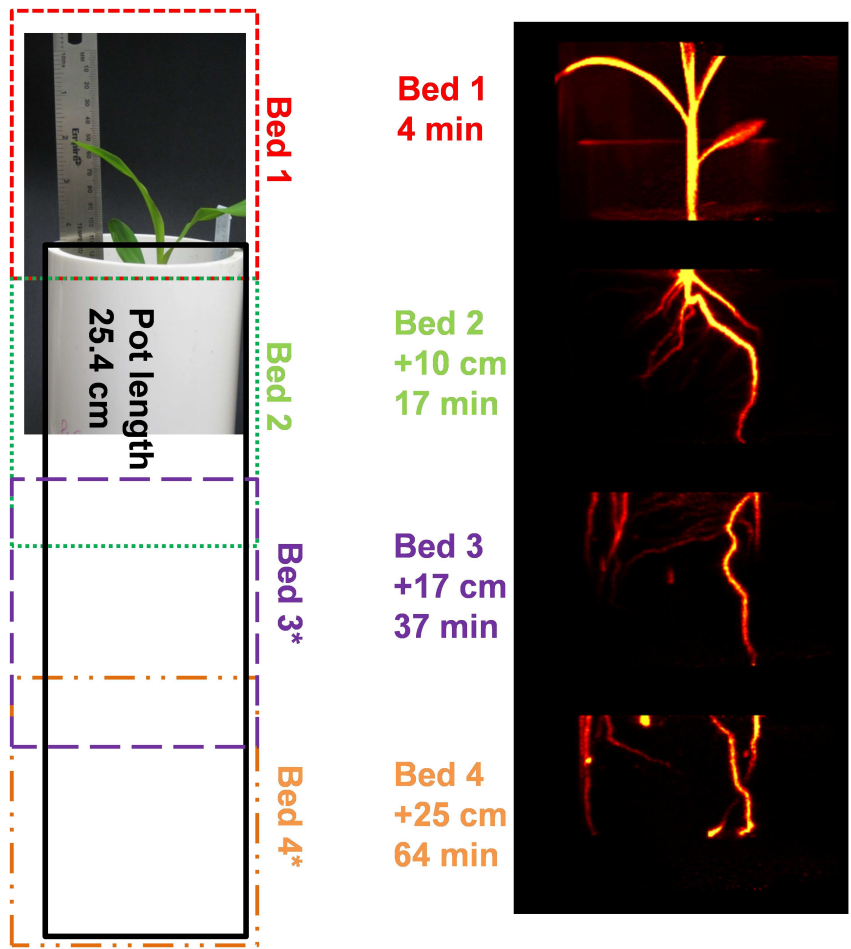}
\caption{Multi-bed scan of a 13-day wild type maize grown in a 10 inch long PVC pot and the segmented PET images of the whole plant, left: the plant and the four bed positions marked with different colored rectangles, right: the PET images of the four different bed positions, the corresponding acquired time points are 4 min, 17 min, 37 min and 64 min as refer to the scan beginning time}
\label{fig:multi-bed}
\end{figure}
The experience from our preliminary plant imaging experiments suggests that bigger trans-axial FOV is very useful. As show in \fref{fig:multi-bed}, the maize roots growth very fast and the roots can go through the 10 inch long tube with in less than two weeks. With the multi-bed scan mode, we can increase the axial FOV, but some fast photoassimilate translocation information will be lost. As the activity is highly accumulated in the small volume of root structure, good image can be reconstructed with around one million of coincidence events which can be acquired within one minute. We can still achieve these 5-minute frames dynamic images using step-and-shot motion in trans-axial direction with 1.25 minute of duration per step. The corresponding trans-axial FOV will close to 16 inch FOV which is four fold of current static system.   

\subsection{System throughput and custom built automated labeling system}

When talking about system throughput, it is not comparable with the conventional optical imaging system already applied to plant phenotyping studies. In fact, one or two minutes of scan can acquire enough coincidence events for reconstructing a good 3D PET image, which is comparable with the CT scanner. Depends on the study purpose, some studies need to continues monitor several hours long physiological process while some studies just need to acquire the PET image at a dedicated time point. One possible high-throughput multi-modality imaging system is to combine the plant PET scanner that work at the snapshot mode with a compact X-ray camera system to image plants with both structure and functional information.

For plant study, the environmental changes during the labeling and imaging stages need to be controlled to avoid the corresponding effects on plant growth. A complete automated plant labeling system that on one hand will improve system's throughput and on the other hand will mostly get rid of the fluctuation of environmental conditions. The plants to be studied are of different size, as a result, it is not easy to build a common labeling system. Some custom made labeling chambers are being or to be built to accommodate more plants in our system. The automatic labeling system also works together with the automatic radio-tracer delivery system so that we can precisely control the radio-labeling process which is very important for a series of repetitive experiments.     

\subsection{Partial volume effect for thin leaves imaging}

\section{Conclusion}
We have developed a dedicated high-resolution plant PET scanner based the detector modules from small-animal PET systems. The scanner features re-configurable system geometry and full control of plant growth environment. The system is composed of two different detector modules with different crystal sizes which shows a reasonable sensitivity of 1.3\% at center of FOV, 18\% system energy resolution, and 1.8 ns time resolution. System spatial resolution is similar to that of commercial microPET$^{\textregistered}$ systems. Phantom studies and preliminary plant imaging experiments show that high quality 3D tomographic and dynamic PET images can be acquired with the full ring configuration. These initial plant imaging studies also clearly demonstrated the functional imaging capability of the plant PET system. Additional studies using N-13, C-11 and other radionuclides are being conduct by collaborating with regional plant scientists. This dedicated plant PET scan possesses a open system geometry, so new high resolution detector modules and optimized geometry configuration can be applied to the system to improve the system performance in terms of spatial resolution, FOV and sensitivity.

\section{Acknowledgments}
We would like to than Dr. Stefan Siegel and Dr. Dongming Hu of Siemens Molecular Imaging, Inc. for technical support in using Siemens  QuickSilver$^{TM}$ electronics, Lee Sobotka, Carmen Dence at Washington University in St. Louis for helpful discussion, and Kinda Abdin, Tom Voller, Lori Strong, Laforest Richard, Patrick Zerkel, Bill Margenau, Greg Gaehle for providing access to radioactive sources. This research was support by U.S. National Science Foundation, grant DBI-1040498. Imaging reconstruction was performed using the facilities of Washington University Center for High Performance Computing, which were partially supported by grant NCRR 1S10RR022984-01A1.

\section*{References}

\bibliographystyle{unsrt}
\bibliography{PlantPET}

\end{document}